\documentclass{iau}
\usepackage{graphicx}

\title[sps13.~~low-mass eclipsing binary] 
{Magnetic activity of several low-mass eclipsing binary}

\author[Liyun Zhang et al.,]   
{Liyun Zhang$^1$ Zhongmu Li$^2$
 Qifeng Pi$^1$ \and Zhongzhong Zhu$^1$ }

\affiliation{$^1$ College of Science/Department of
Physics, Guizhou
University, Guiyang 550025, China \\ email: {\tt Liy\_zhang@hotmail.com} \\[\affilskip]
$^2$Institute for Astronomy and History of Science and Technology, Dali University, Dali, China}

\pubyear{2012}
\volume{xxx}  
\pagerange{119--126}
\jname{High-precision tests of stellar physics from high-precision photometry}
\editors{D. Soderblom eds.}
\begin{document}

\maketitle

\begin{abstract}
We present our new photometric and spectroscopic observations of NSVS 02500276, NSVS 07453183, NSVS 11868841, NSVS 06550671 and NSVS 10653195. The first flare-like event was detected on NSVS07453183. Using the Wilson-Devinney program, the preliminary orbital solutions and starspot parameters are derived. The chromospheric activity indicators show NSVS10653195 and NSVS06550671 are active. Then, we discuss the starspot evolution on the short and long term scale. In the end, we give our future plan.

\end{abstract}

\firstsection 
\section{Introduction}
 Although M-type stars are the most populous stellar objects in the Galaxy, the physical properties (magnetic activities) are poorly understood  (Ribas 2006, 2007). Until now, only about a dozen low-mass eclipsing binaries in detached systems have been detected and studies in detail. Even in these small numbers, the detailed comparison of the observed masses and radii with theoretical predictions has revealed large disagreements. The reason might be magnetic activity (Morales et al. 2010), so it is necessary to monitor.
\section{light curve and spectral analysis}
\begin{figure*}
\centering
\includegraphics[width=2.5cm,height=1.8cm]{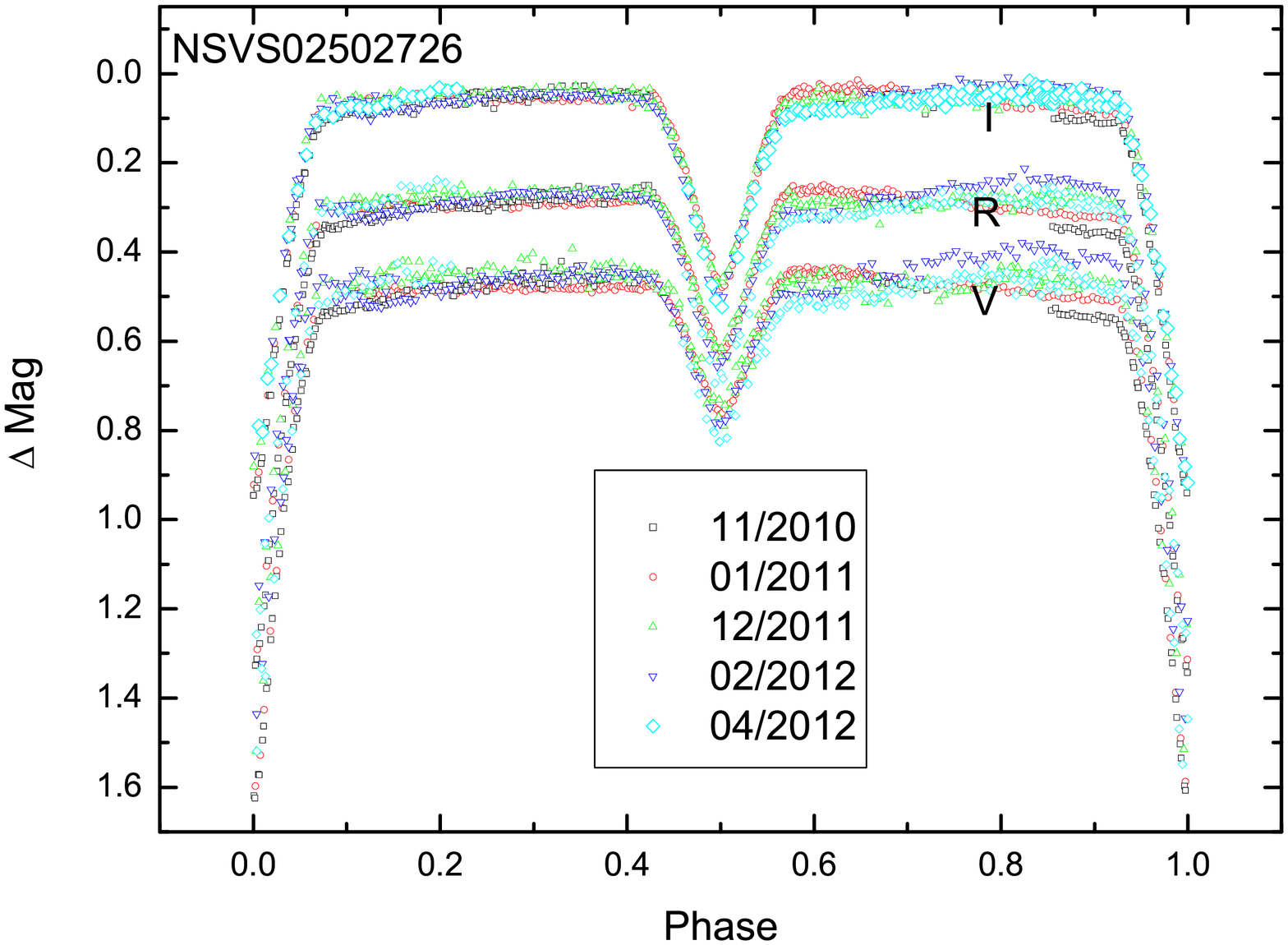}
\includegraphics[width=2.5cm,height=1.8cm]{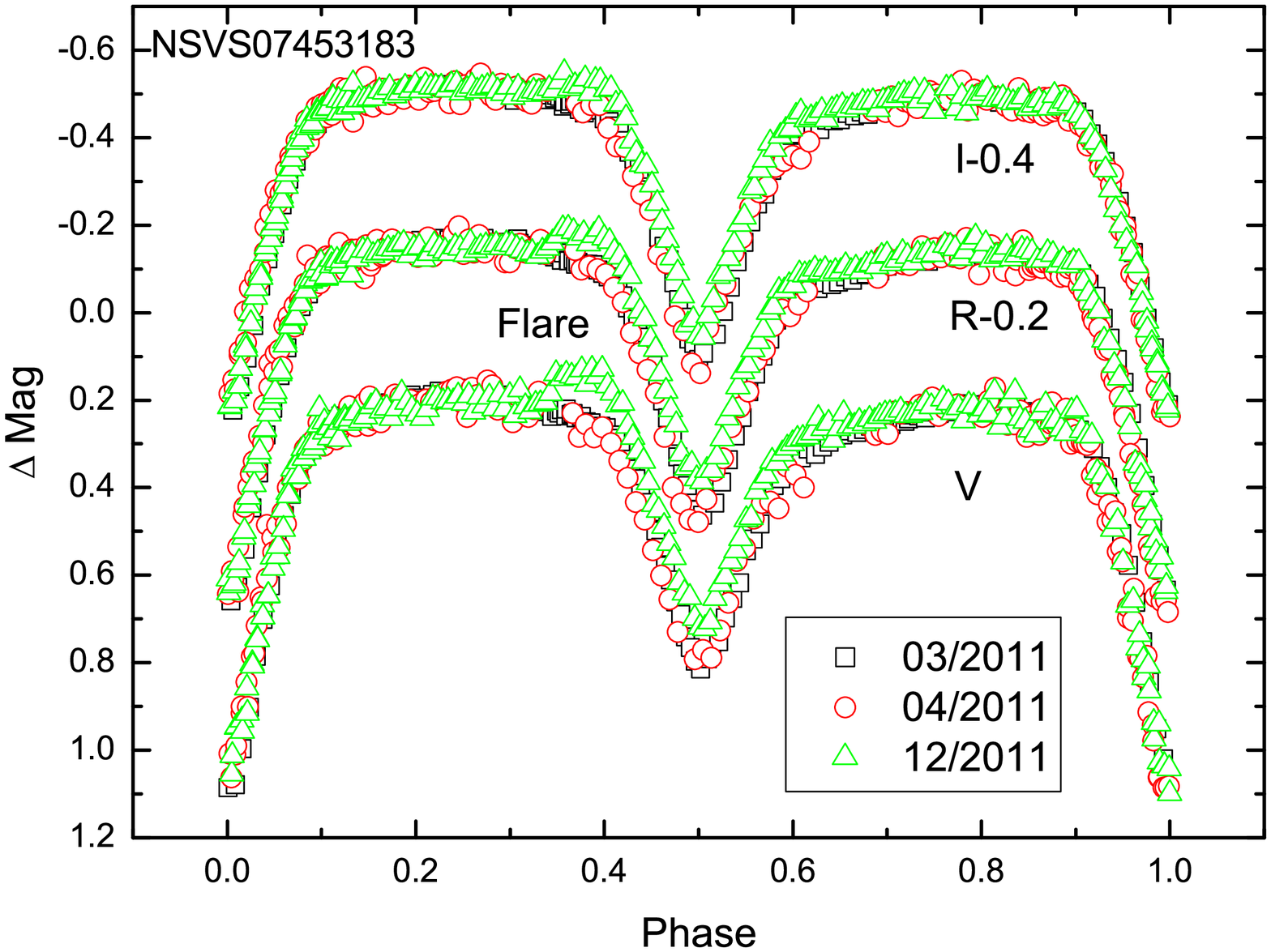}
\includegraphics[width=2.5cm,height=1.8cm]{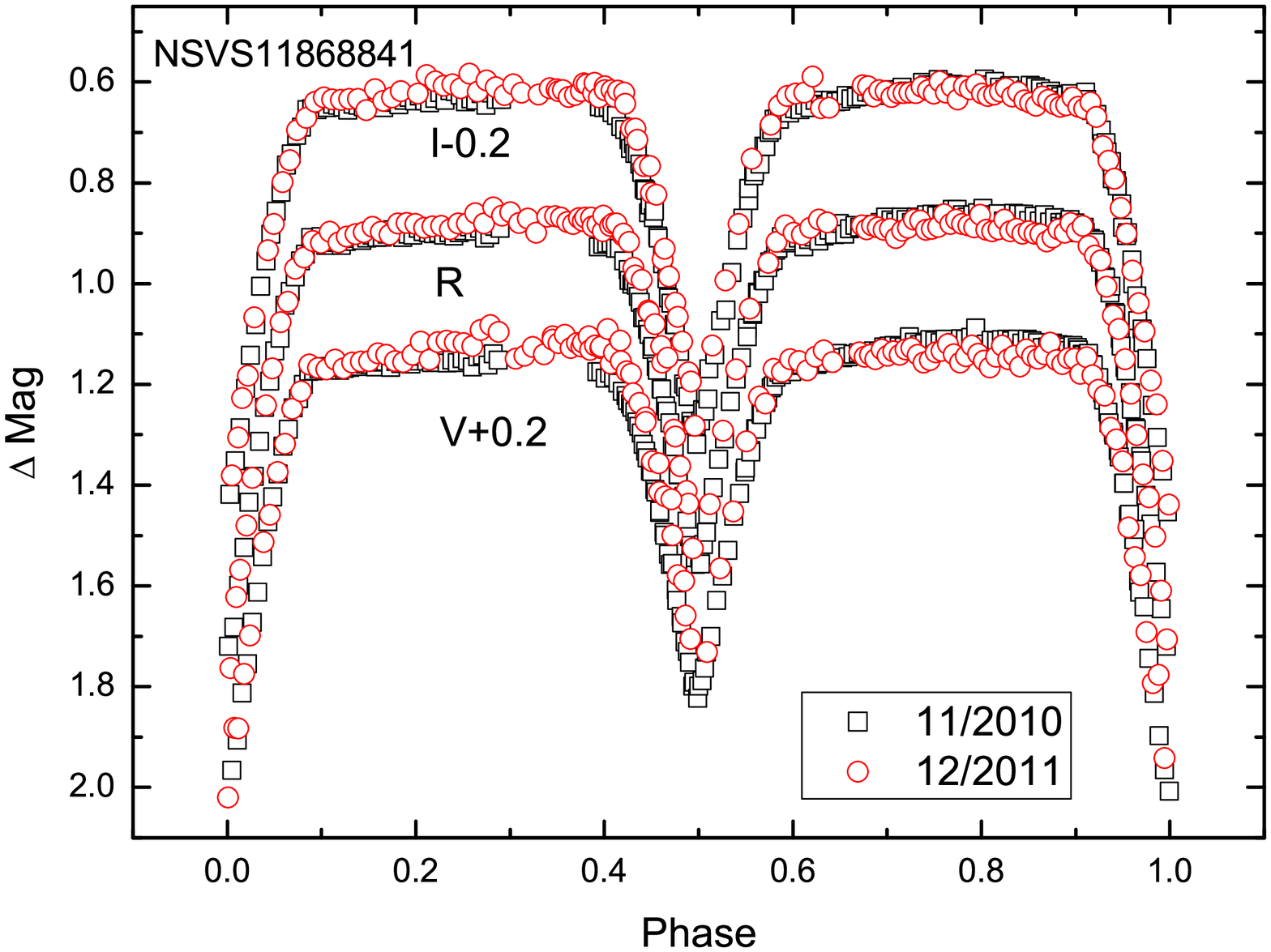}
\includegraphics[width=2.5cm,height=1.8cm]{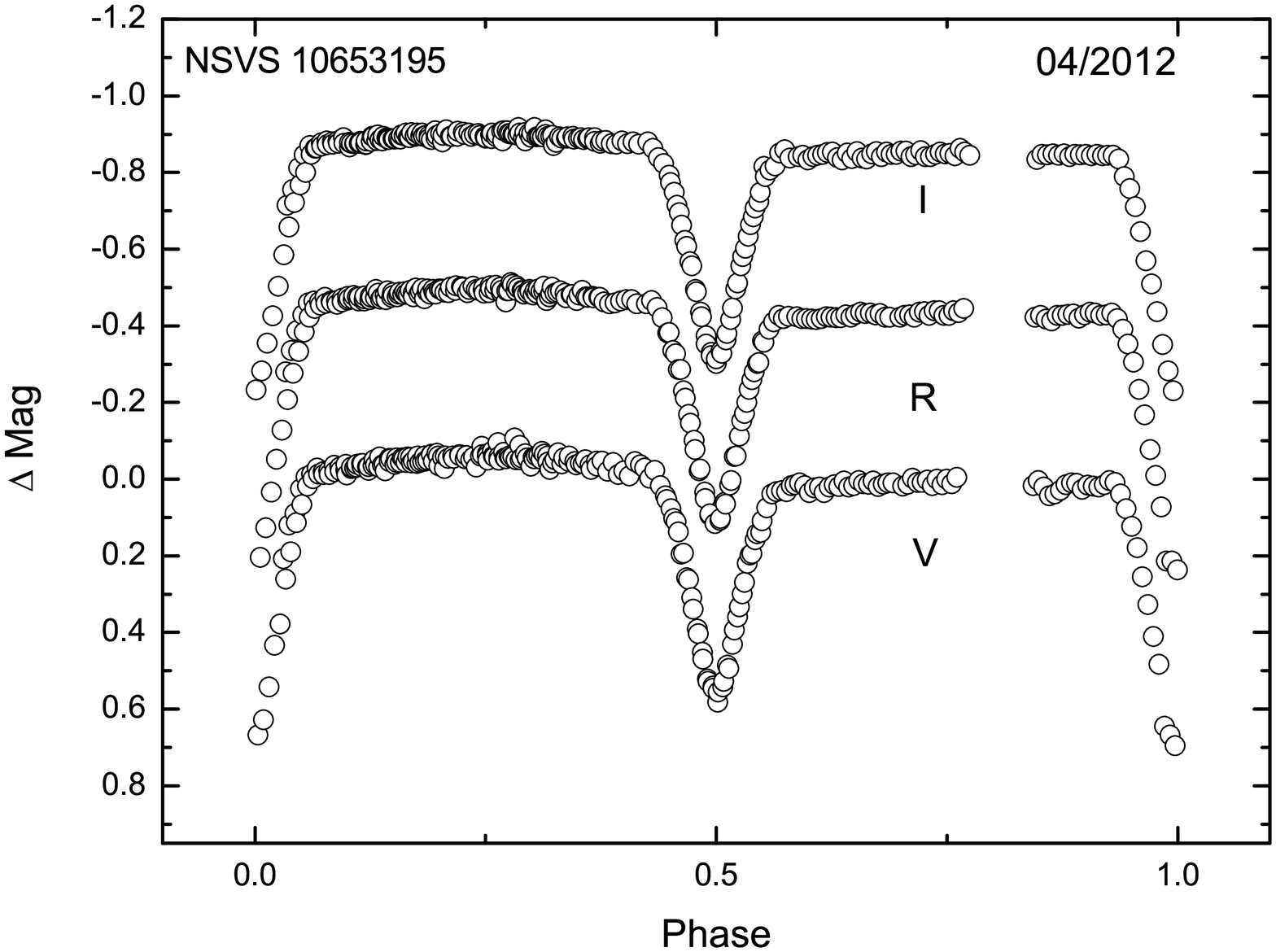}
\includegraphics[width=2.5cm,height=1.8cm]{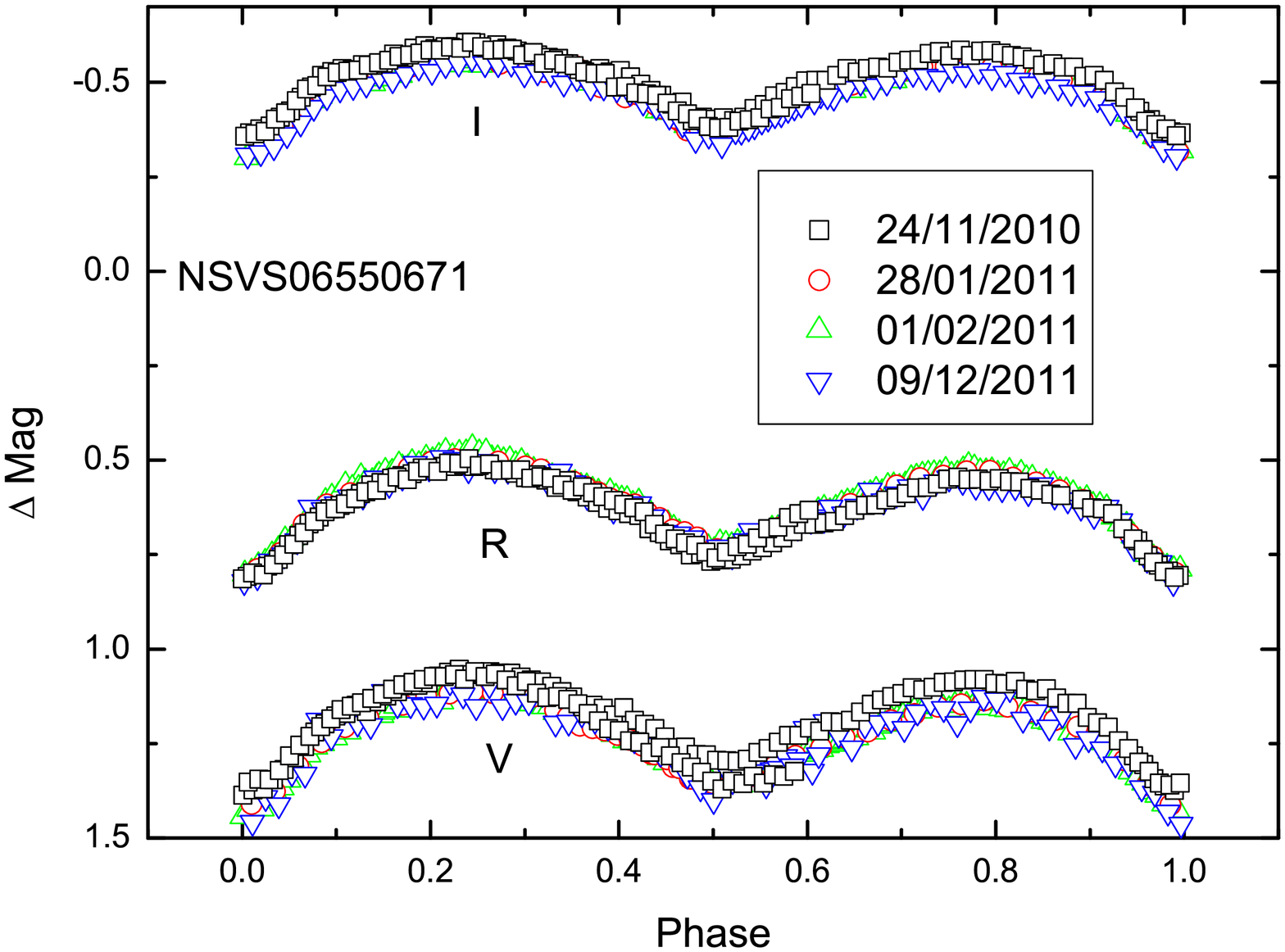}
\caption{V, R, and I observations of the five low-mass eclipsing at Xinglong
station of NAOC.}
\end{figure*}
Our new CCD photometric observations of the five low-mass eclipsing binary ( NSVS 02500276, NSVS 07453183, NSVS 11868841, NSVS 06550671 and NSVS 10653195) were carried out from 2010 to 2012 with an 85-cm telescope (Zhou et al. 2009) at the Xinglong station of the National Astronomical Observatories of China (NAOC). The VRI light curves are displayed in Figure.1. Our spectroscopic observations of the NSVS10653195 and NSVS06550671 (Fig. 2) were carried out with the OMR spectrograph (centered at about 4280 ${\AA}$ with a reciprocal dispersion of 1.03 ${\AA}$/pixel) of the 2.16m telescope at the Xinglong station of NAOC.\\
\begin{figure*}
\centering
\includegraphics[width=2.1cm,height=1.5cm]{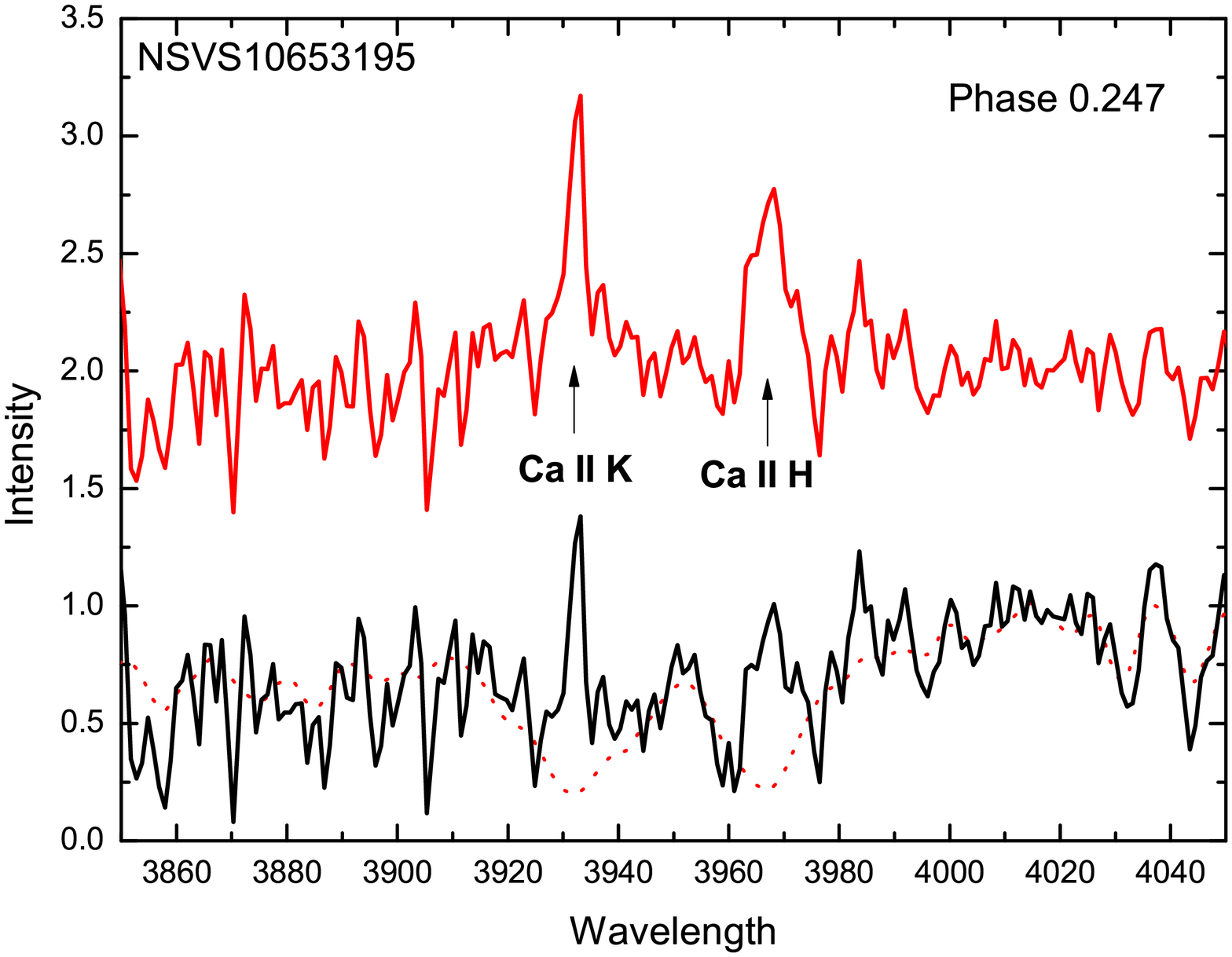}
\includegraphics[width=2.1cm,height=1.5cm]{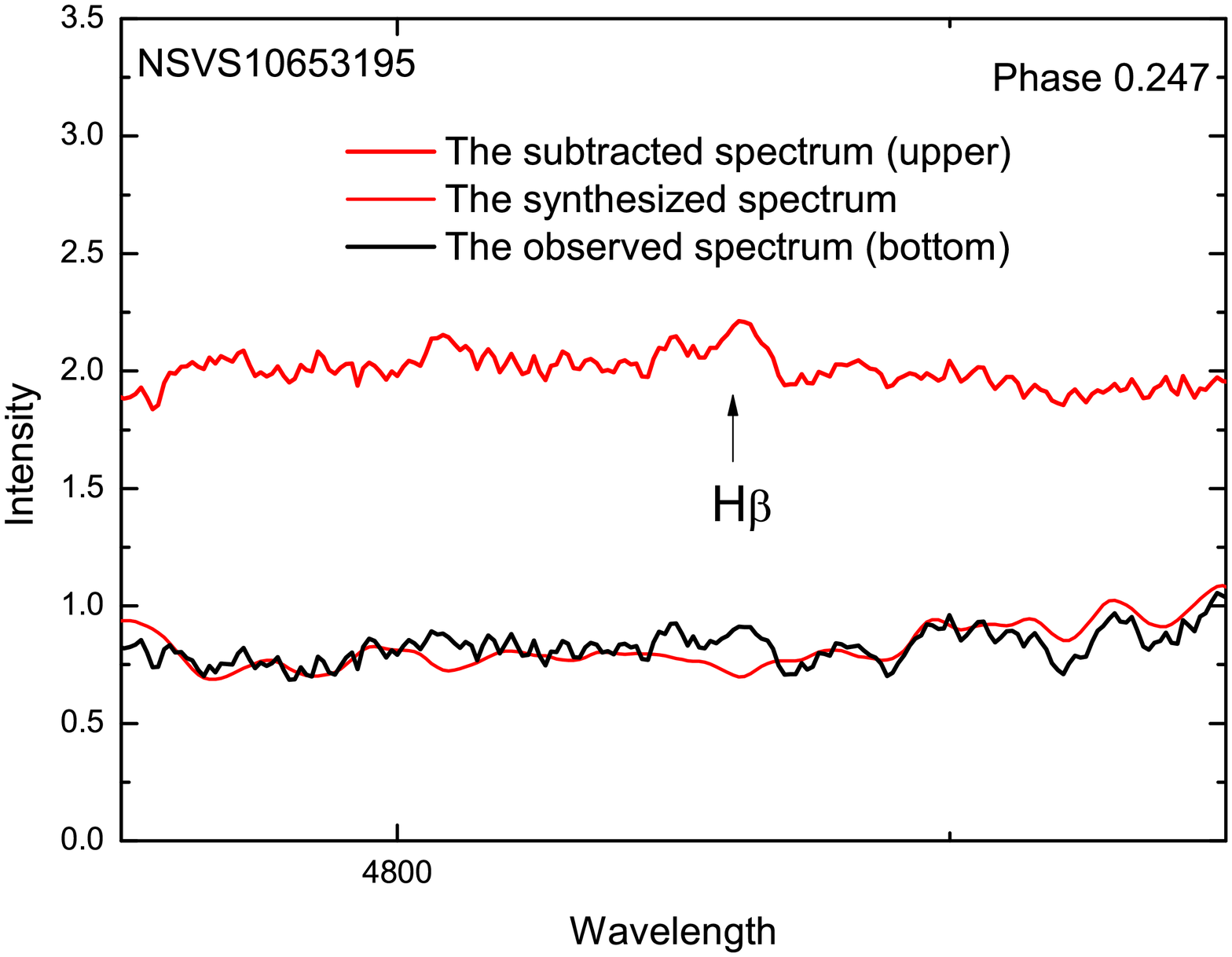}
\includegraphics[width=2.1cm,height=1.5cm]{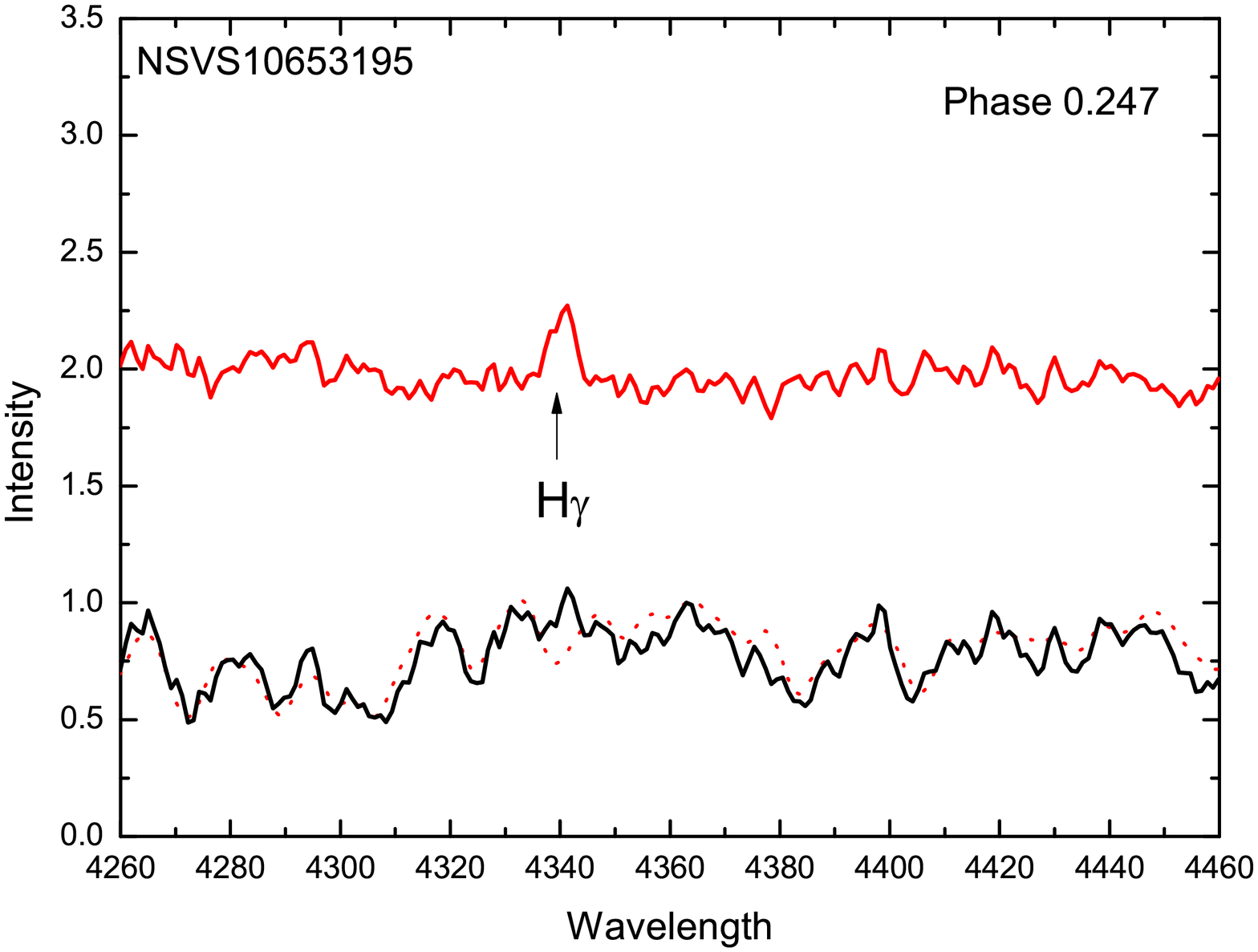}
\includegraphics[width=2.1cm,height=1.5cm]{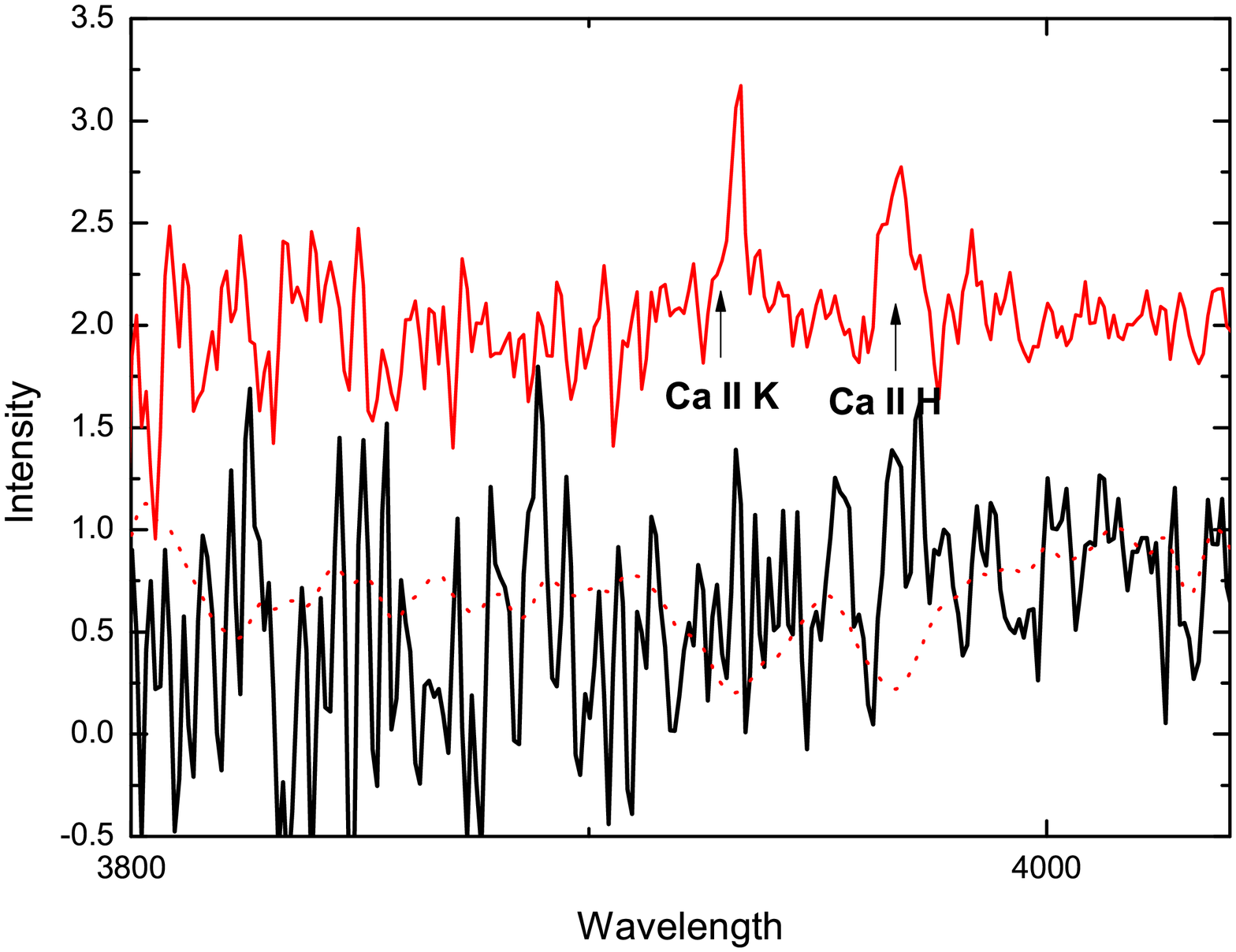}
\includegraphics[width=2.1cm,height=1.5cm]{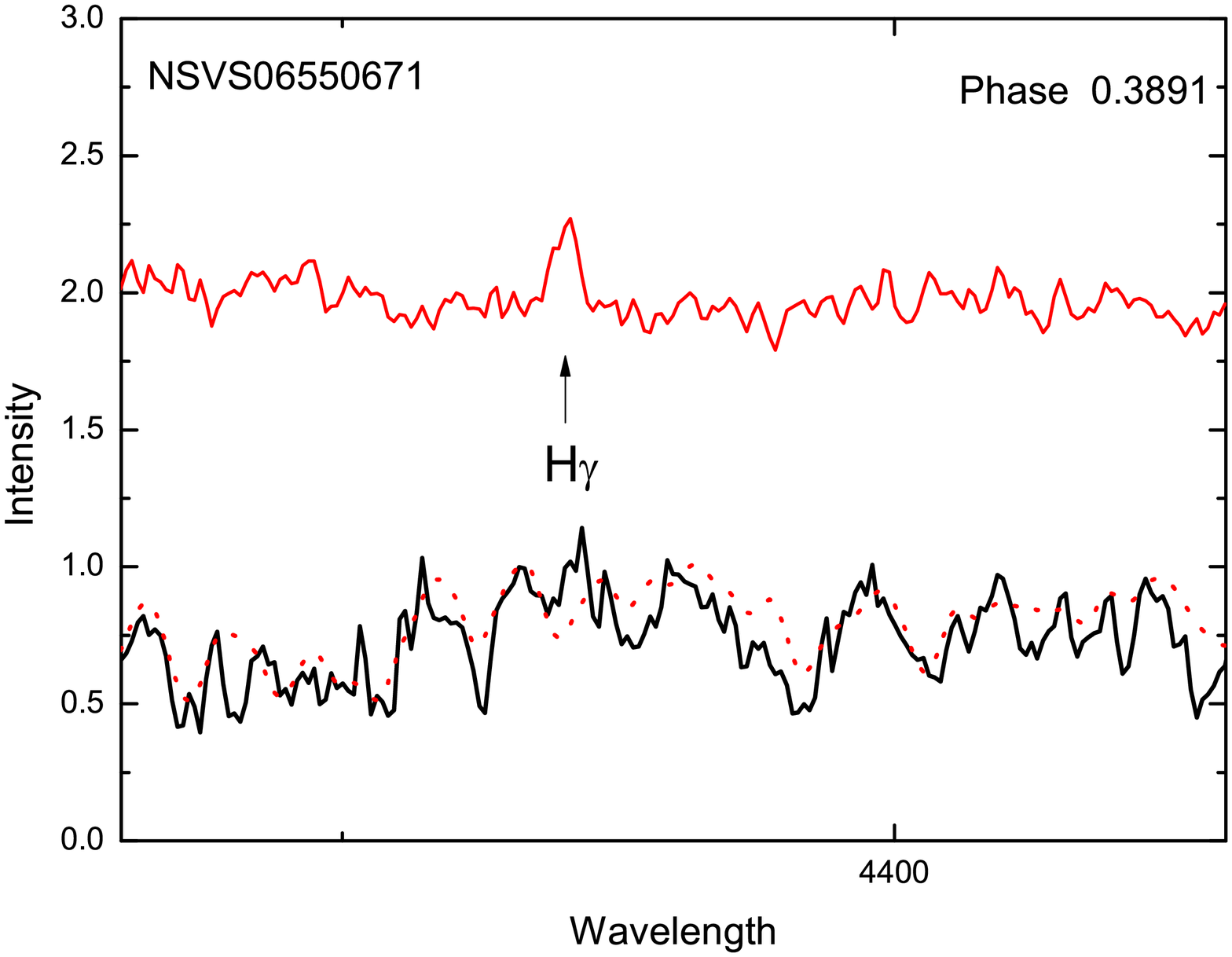}
\includegraphics[width=2.1cm,height=1.5cm]{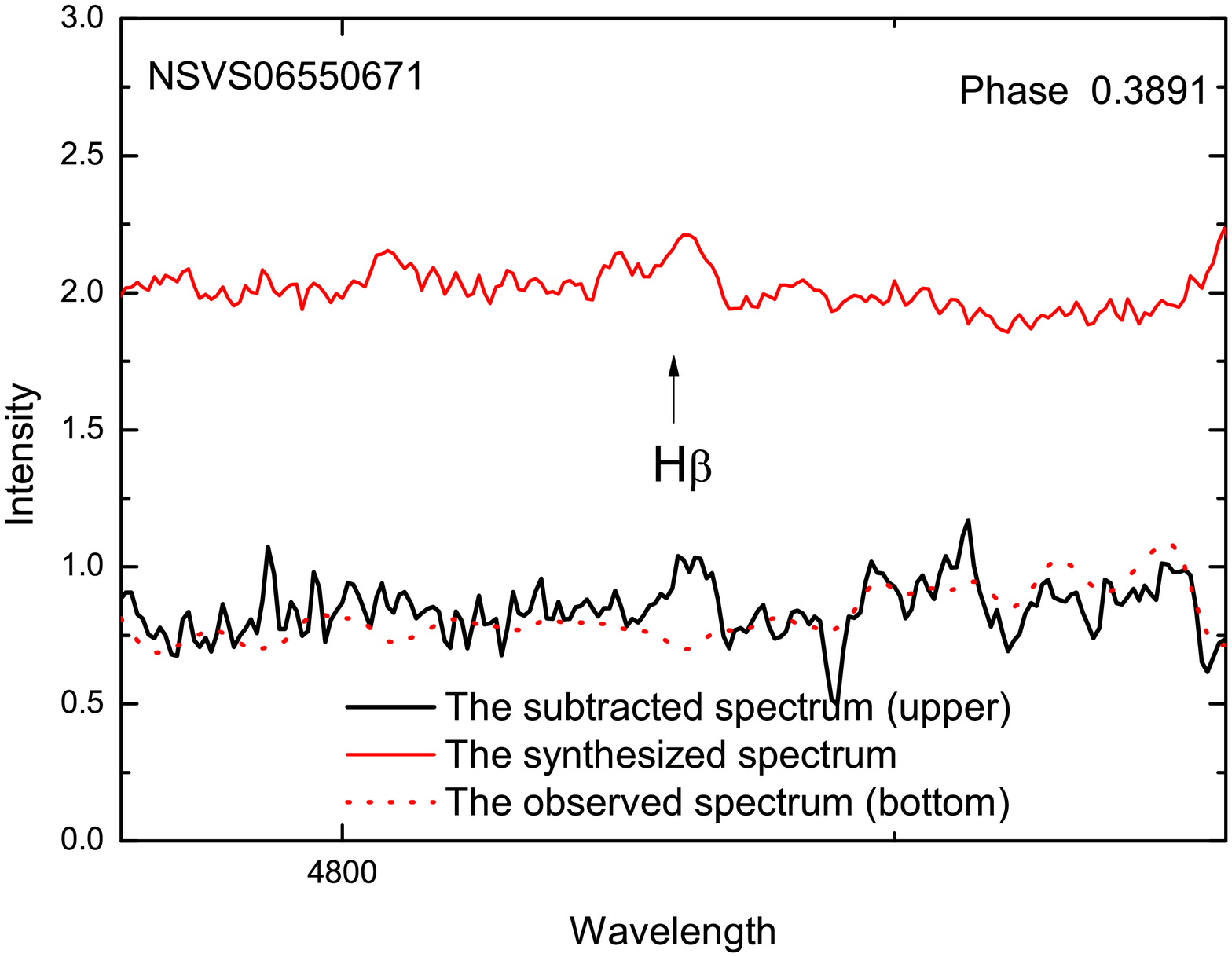}
\caption{The observed, synthesized, and subtracted
spectra for the $\mbox{Ca~{\sc ii}}$ H\&K, H$_{\beta}$ and H$_{\gamma}$ lines. The dotted
lines represent the synthesized spectra and the upper are
the subtracted one.}
\end{figure*}
\begin{figure}
\centering
\includegraphics[width=3.1cm,height=1.8cm]{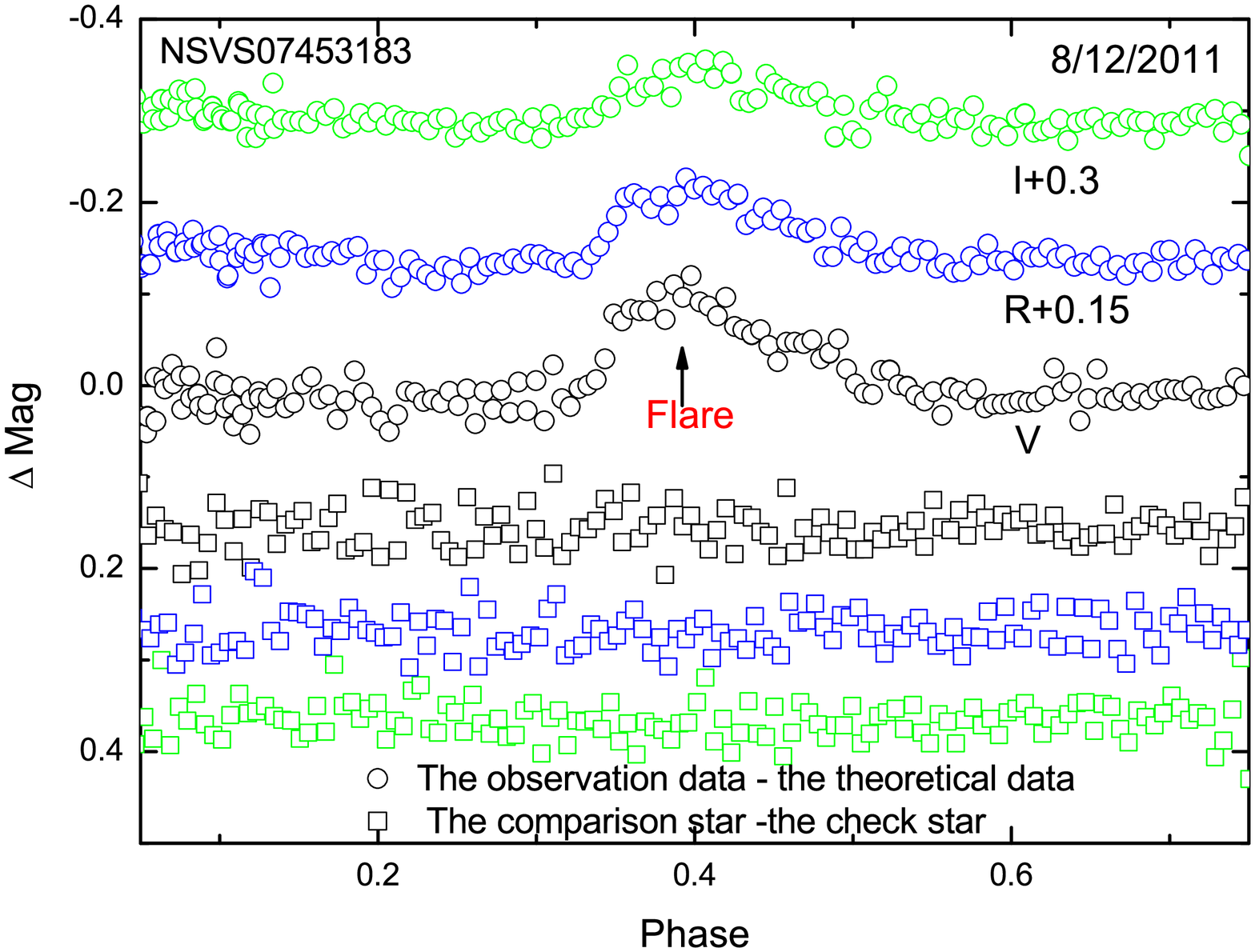}
\includegraphics[width=3.1cm,height=1.8cm]{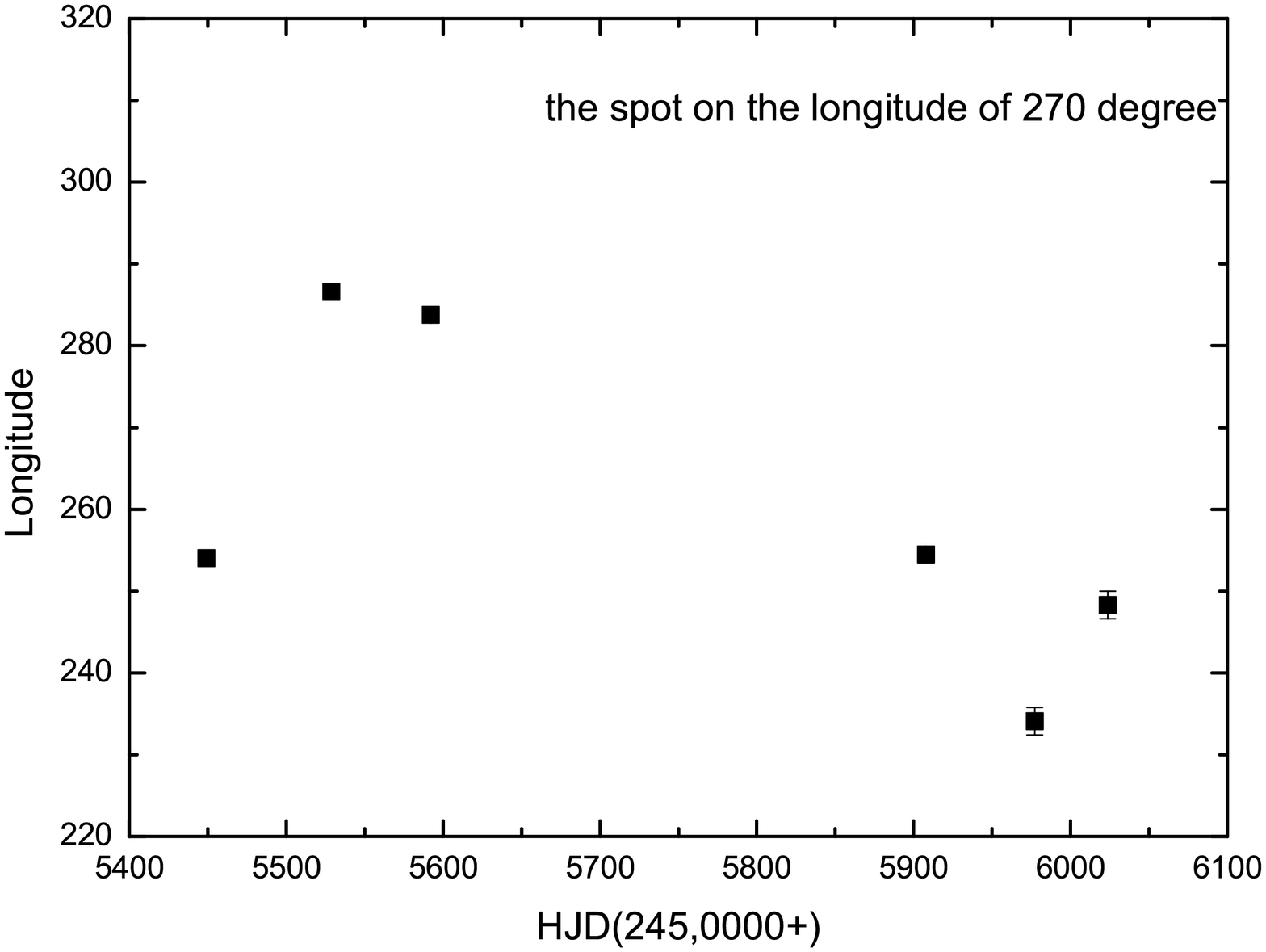}
\includegraphics[width=3.1cm,height=1.8cm]{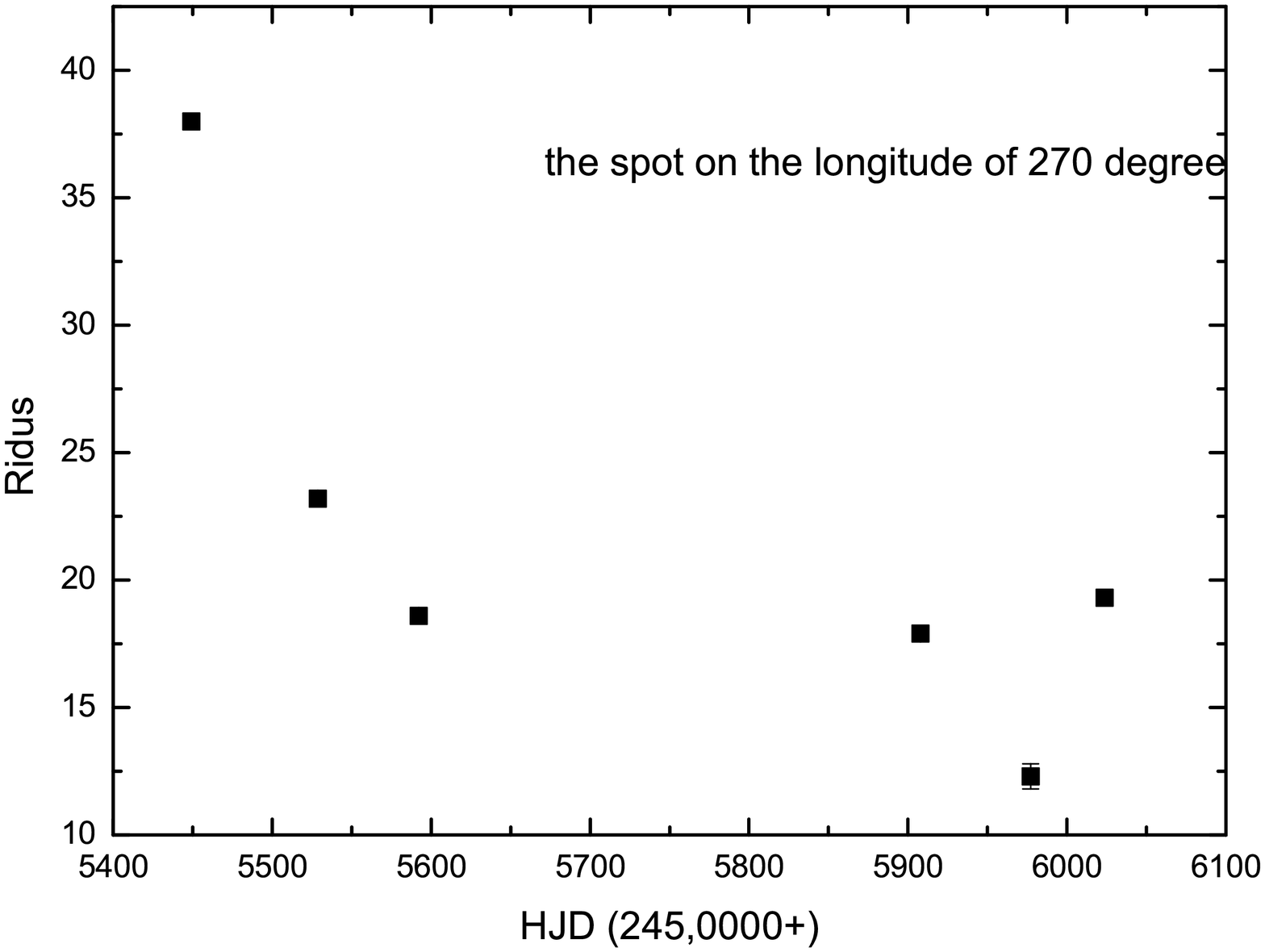}
\caption{The flare of NSVS 07453183 in V, R, and I bands (left). The spot longitude (middle) and radius (right) evolution of NSVS02502726 on a short and long term scale.}
\end{figure}
\indent For NSVS02502726, NSVS 11868841 and NSVS 06550671, our VRI light curves and the previous published radial velocities ($\c{C}akirli$ et al. 2009; $\c{C}akirli$ et al. 2010; Dimitrov \& Kjurkchieva 2010)  were analyzed simultaneously using the updated Wilson-Devinney code to revise the photometric-spectroscopic elements (Wilson \& Devinney 1971; etc).  For NSVS10653195 and NSVS07453183, we only obtained the photometric orbital parameters. The preliminary orbital parameters are taken from the prior results (Coughlin et al. 2007; Wolf et al. 2009). The spectra were analyze by the spectral subtraction technique using the program STARMOD (Barden1985; Montes et al.1997). The synthesized spectra is constructed from artificially rotationally broadened, radial-velocity shifted, and weighted spectra of two inactive stars with the same spectral type and luminosity class.\\
\section{Conclusion and perspective}
The results can summarized as follows. 1. The orbital and starspot parameters were obtained. 2. The first flare-like event was detected on NSVS07453183 at phase 0.39 with a maximum amplitude of 0.1 mag in the V, 0.076 in R and 0.05 in I band (Fig.3 left). 3. The longitude and radius of the spot around 270 degree of NSVS 02502726 change on a time scale of two months (Fig.3 midlle and right). 4. The Ca II H \& K, H$_{\beta}$ and H$_{\gamma}$ lines show that NSVS10653195 and NSVS06550671 are active. We propose: 1. Chromospheric activity of low mass stars using Lijiang 2.4m, Xinglong 2.16m and Gou Shou Jing telescope, NAOC. 2. Photometric monitor to study magnetic active cycle.
\begin{acknowledgements}
We are very grateful to Dr. Montes D., Gu S. H., Han J. l., Zhou A. Y., Zhou X., Jiang X. J., and Fang X. S. The
work is supported by the NSFC under grant No. 10978010, 11263001, 11203005 and 10373023. This work was partially Supported by the Open Project Program of the Key Laboratory of Optical Astronomy, NAOC, CAS.
\end{acknowledgements}

\end{document}